\documentclass[aps,preprint]{revtex4}
\usepackage{graphicx} 
\usepackage{amsmath,amsfonts,amssymb}

\newcommand{\ra}{\rightarrow}
\newcommand{\eps}{\epsilon}

\newcommand{\CF}{{\mathcal F}}

\newcommand{\bp}{{\mathbf p}}

\newcommand{\bj}{{\mathbf j}}
\newcommand{\bx}{{\mathbf x}}
\newcommand{\by}{{\mathbf y}}

\newcommand{\bE}{{\mathbf E}}
\newcommand{\bG}{{\mathbf G}}
\newcommand{\bK}{{\mathbf K}}
\newcommand{\be}{{\mathbf e}}

\newcommand{\Dp}{{\Delta_{\Omega}}}

\begin{document}

\author{Anton Kapustin}

\affiliation{California Institute of Technology, Pasadena, CA 91125, U.S.A.}

\title{A soluble model of a Non-Equilibrium Steady State: the van Kampen objection and other lessons}

\begin{abstract}
    A simple model of charge transport is provided by a classical particle in a smooth random potential and a dissipative coupling to the environment in the form of Markovian noise and friction. The corresponding Non-Equilibrium Steady State (NESS) can be determined analytically when both the disorder and dissipation are weak. We use it to illuminate some foundational issues in non-equilibrium statistical mechanics. We show that Linear Response Theory has a nonempty regime of validity only in the presence of a dissipative coupling to the environment,  thereby validating van Kampen's objection. We also show that the Principle of Minimum Entropy Production does not determine the NESS beyond linear order in the electric field, while entropy maximization fails to produce the correct NESS already at linear order.
\end{abstract}

\maketitle

\section{Introduction}

Linear response theory (LRT) is widely used, but it has subtle points which are not always appreciated. For example, while Kubo formulas for electric conductivity, thermal conductivity, viscosity, etc., are supposed to describe properties of non-equilibrium steady states (NESS) which can exist only in open systems, derivations of Kubo formulas assume a closed system. This point has been forcefully expressed by N. van Kampen \cite{vankampenobjection} who argued that a coupling to the environment is needed to ensure macroscopic linearity of response. The usual answer is that despite van Kampen's objection, if the coupling to the environment is sufficiently weak, the system can be treated as closed. This, however, raises the question about the precise meaning of "sufficiently weak", that is, about the size of  corrections to the Kubo formulas, both linear and nonlinear. One also wonders about the nature of the NESS in the nonlinear regime and whether anything can be said about it without a detailed knowledge of the coupling to the environment. It has been proposed, for example, that the Minimum Entropy Production (MINEP) principle can be used to determine NESS \cite{Prigogine,deGrootMazur,KleinMeijer}, just like the Maximum Entropy (MAXENT) principle determines the Gibbs distribution for equilibrium states \cite{Jaynes}. It is difficult to test this, however, since there are very few situations where a NESS is known in the non-linear regime.

In this note we study a simple but physically reasonable model of electric conduction where such questions can be addressed. It is a version of the classical Drude model where the potential is assumed to be smooth, weak, and random. We also add a dissipative coupling to the environment in the form of a Markovian noise and a velocity-dependent friction force which are related by the fluctuation-dissipation theorem. Such dissipative  couplings are necessary for a NESS to exist beyond linear order in the electric field (see below). One can also interpret the dissipative coupling as describing inelastic scattering off phonons. 

The model can be analyzed analytically in the weakly non-linear regime, i.e. small electric fields. The most interesting case is when the dissipation is very weak, so that the elastic scattering dominates over the inelastic one. In this regime, one can compute conductivity in two ways. First, one can neglect dissipation and use the Kubo formula. Second, one can determine NESS at a small but nonzero dissipation and compute the resulting current. The two methods give results which differ by a quantity of  order $O(\nu)$, where $\nu$ is the dissipative coupling. If the inelastic resistivity is much lower than the elastic one, one can simply neglect inelastic scattering. We also evaluate the $O(\nu)$ correction to the conductivity and find that the Matthiessen rule is violated. 

In the second approach, one cannot set the dissipative coupling to zero, because corrections to the NESS which are quadratic in the electric field $E$ diverge in this limit. If one takes $\nu$ to zero, the region of applicability of LRT shrinks to nothing. LRT is valid provided Joule heating (the relative increase of the average kinetic energy due to a nonlinear correction to the distribution function) is negligible. This is true if $E$ is smaller than a quantity of order $\sqrt\nu$. This point was recently discussed in Ref. \cite{GloriosoHartnoll} in a qualitative way. 

An explicit formula for NESS allows us to test some popular variational principles proposed to determine the form of NESS, such MAXENT and MINEP. We show that MAXENT fails already at linear order in $E$, while MINEP is satisfied at linear order but fails at quadratic order.  

The author would like to thank Gregory Falkovich and Boris Spivak for numerous illuminating discussions and to Weizmann Institute of Science for hospitality. This work was supported in part by the U.S.\ Department of Energy, Office of Science, Office of High Energy Physics, under Award Number DE-SC0011632 and by the Simons Investigator Award.



\section{Newtonian particle in a weak random potential}

We consider a classical particle in $d$ spatial dimensions subject to a potential $-{\bE}\cdot {\bx}+U({\bx})$, where $U({\bx})$ is smooth and random. The kinetic energy is  $\eps(p)=p^2/2$, i.e. we use the units in which the particle's mass is $1$. Then we can identify velocity and momentum. Let $\hat\bp$ be the unit vector in the direction of $\bp$.
The distribution function in phase space $\CF(t,\bx,\bp)$ obeys the Liouville equation
\begin{equation}\label{eq:Liou}
    \frac{\partial \CF}{\partial t}+\frac{\partial}{\partial x_i}\left(p_i \CF\right)+\frac{\partial}{\partial p_i}\left((E_i-\partial_i U) \CF\right)=0.
\end{equation}
One would like to obtain a closed equation for the distribution function averaged over the realizations of the potential $U(\bx)$. This is possible to do in the van Hove limit \cite{VH1954}, where we rescale $t\mapsto t/\eps^2$, $U\mapsto \eps U$, $\bE\mapsto \eps^2 \bE$ and take $\eps$ to be vanishingly small \cite{vanKampenstochastic,KestenP}. The scaling of $t$ and $U$ mean that we take the random potential to be weak and simultaneously look at the long-time behavior. The chosen scaling of $\bE$ means that the change of energy between collisions due to electric field is parametrically small. (However, as discussed below, the change of energy between {\it effective collisions} can be of order $1$.)

The disorder-averaged distribution function $f(t,\bp)=\langle\CF(t,\bx,\bp)\rangle$ is independent of $\bx$ and satisfies a Fokker-Planck equation \cite{KestenP}:
\begin{equation}\label{eq:FPnodiss}
\frac{\partial f}{\partial t}+E_i\frac{\partial f}{\partial p_i}=\frac{W}{p^3}\Dp f,
\end{equation}
where $p=|\bp|$, $\Dp$ is the Laplacian on the unit sphere in the $\bp$-space, and $W$ is a positive constant determined by the 2-point function of $U(\bx)$:
\begin{equation}
    W=-\int_0^\infty \frac{d}{dr}\langle U(0) U(r)\rangle \frac{dr}{r}.
\end{equation}
A derivation of eq. (\ref{eq:FPnodiss}) along the lines of Ref. \cite{KestenP} is given in the Appendix.

The r.h.s. of eq. (\ref{eq:FPnodiss}) describes angular diffusion in momentum space. Since the scattering is elastic, there is no diffusion in kinetic energy. In fact, for $\bE=0$ the momentum-space average of an arbitrary function of $p=|\bp|$ is an integral of motion: $\frac{d}{dt}\int h(p) d^dp=0$.
This is no longer true for $\bE\neq 0$. In particular, for the average kinetic energy we get
\begin{equation}
    \frac{d}{dt}\int\frac{\bp^2}{2} d^d p=\bE\cdot \int  \bp\, d^d p .
\end{equation}
Thus, for $\bE\neq 0$ a stationary distribution $f$ is impossible unless the average current vanishes. In the next section we will rectify this by adding dissipation.

The entropy $S=-\int f \log f d^dp $ satisfies an H-theorem:
\begin{equation}\label{eq:dsdt}
\frac{dS}{dt}=\int \frac{W P_{jk}}{p}\frac{\partial_j f\,\partial_k f}{f} d^dp  \geq 0.
\end{equation}
Here $P_{jk}=\delta_{jk}-p_jp_k/p^2$ is the transverse projector in momentum space. $dS/dt$ vanishes if and only if $f$ is a function of $p$ only. This implies that for $\bE\neq 0$ eq. (\ref{eq:FPnodiss}) has no stationary solutions at all, while for $\bE=0$ the most general smooth stationary solution has the form $f(\bp)=f_0(p)$ for some non-negative  function $f_0(p)$. The equation does not fix the form of $f_0(p)$. 

Although there is no stationary state for $\bE\neq 0$, one can formally find a linear response to an infinitesimal time-independent $\bE$ by linearizing Eq. (\ref{eq:FPnodiss}). Writing $f=f_0+f_1+O(E^2)$, where $f_0$ is independent of $\bE$ and $f_1$ is linear in $\bE$, and dropping terms of order $E^2$, one finds an equation for $f_1$:
\begin{equation}
    \Dp f_1=\frac{p^3}{W} (\bE\cdot \hat\bp) f_0'(p),
\end{equation}
which is solved by 
\begin{equation}\label{eq:f1nondiss}
    f_1=-\frac{1}{2W}(\bE\cdot \hat\bp)p^3 f_0'(p).
\end{equation} 
Setting $d=3$ and evaluating the average current $\bj=\int \bp f d^3p$ using the Maxwell-Boltzmann distribution for $f_0$ we find $\bj=\sigma \bE$, where 
\begin{equation}\label{eq:sigmalinear}
    \sigma=\frac{16 T^{3/2}}{W{\sqrt {2\pi}}}.
\end{equation}

One can get the same result from the Kubo formula which does not require one to determine the NESS. In this approach, one computes the diffusion coefficient:
\begin{equation}
D_T=\frac{1}{3}\int_0^\infty g_T(t) dt,
\end{equation}
where $g_T(t)=\langle \bp(t)\cdot\bp(0)\rangle$. The momentum correlator $g_T(t)$ involves averaging over the random potential as well as the initial conditions of the particle's trajectory. It is convenient to average over the initial momentum $\bp_0$ last. The correlator at a fixed $\bp_0$ is given by
\begin{equation}
    g_{\bp_0}(t)=\bp_0\cdot \int \bp K(t,\bp;\bp_0) d^3 p,
\end{equation} 
where $K(t,\bp;\bp_0)$ is the solution of (\ref{eq:two}) with the initial condition $f(0,\bp)=\delta^3(\bp-\bp_0)$. It is given by
\begin{equation}
K(t,\bp;\bp_0)=\frac{\delta(p-p_0)}{4\pi p_0^2}\sum_{l=0}^\infty e^{-W l(l+1) t/p_0^3} (2l+1) P_l(\cos\theta),
\end{equation}
where $\theta$ is the angle between $\bp$ and $\bp_0$ and $P_l(x)$ is the Legendre polynomial of degree $l$. Then
\begin{equation}
    g_{\bp_0}(t)=p_0^2 e^{-2Wt/p_0^3},
\end{equation}
and the corresponding diffusion coefficient is $D_{\bp_0}=\frac{p_0^5}{6W}.$
$D_T$ is obtained by averaging $D_{\bp_0}$ over $\bp_0$ using the Maxwell-Boltzmann distribution:
\begin{equation}
    D_T=\frac{16T^{5/2}}{W\sqrt {2\pi}}.
\end{equation}
The Einstein relation $D_T=T\sigma$ then gives (\ref{eq:sigmalinear}).

Comparing (\ref{eq:sigmalinear}) with Drude theory, we see that the Drude mean free time is $\tau=16 T^{3/2}/W\sqrt {2\pi}$. This can be explained as follows. If the typical potential height is $u$ and the typical distance between impurities is $a$, then $W\sim u^2/a$. Since the   velocity of the particle $\sim\sqrt T$, a typical scattering angle in each collision is of order $u/T$ which is small in the van Hove limit. It takes of order $(T/u)^2$ random collisions to change the momentum direction substantially. The corresponding time is of order $T^{3/2}a/u^2$. 

Note that the mean free path is of order $T^2/W$, and the work done by the electric field over the mean free path can be comparable to the average kinetic energy $3T/2$ in the van Hove limit. Naively, LRT should be applicable when this work is much smaller than $T$, i.e. when $E\ll W/T$. To check this, one might try to continue the expansion of $f$  to quadratic order in $\bE$, $f=f_0+f_1+f_2+O(E^3).$ But one runs into a problem: the equation for $f_2$ has no solution. This must be so because eq.  (\ref{eq:FPnodiss}) does not have any stationary solutions for $\bE\neq 0$. To fix this issue, we need to add dissipation. 

\section{Adding dissipation}

To add dissipation (i.e. a coupling to the environment), we include a friction force $-\nu \bp$ as well as a Gaussian delta-correlated noise. This is equivalent to replacing  the Liouville equation (\ref{eq:Liou}) with a Fokker-Planck equation
\begin{equation}
\frac{\partial \CF}{\partial t}+\frac{\partial}{\partial x_i}\left(p_i \CF\right)+\frac{\partial}{\partial p_i}\left((E_i-\partial_i U-\nu p_i) \CF\right)=\nu T\frac{\partial^2\CF}{\partial p_i\partial p_i}.
\end{equation}
The coefficient of the diffusion term on the r.h.s. is fixed by the requirement that for $\bE=0$ the Maxwell-Boltzmann distribution $e^{-(U+p^2/2)/T}$ is a stationary solution of the Fokker-Planck equation. This is the fluctuation-dissipation relation.

Averaging over the random potential in the van Hove limit and assuming $\nu$ is of order $\eps^2$, we get an equation for the disorder-averaged distribution function (see Appendix for details):
\begin{equation}\label{eq:two}
    \frac{\partial f}{\partial t}+\frac{\partial}{\partial p_i}\left((E_i-\nu p_i)f\right)=\frac{W}{p^3}\Dp f+\nu T\Delta f,
\end{equation}
where $\Delta$ is the Laplace operator in momentum space. One can show that for $\bE=0$ eq. (\ref{eq:two}) has a unique time-independent solution: the Maxwell-Boltzmann distribution $f_0(\bp)\sim e^{-p^2/2T}$. To find NESS solutions for $\bE\neq 0$ it is convenient to introduce a dimensionless momentum variable $\by=\bp/\sqrt T$ and two dimensionless parameters 
\begin{equation}
    {\bG}=\frac{{\bE}}{\nu\sqrt T},\quad B=\frac{W}{\nu T^{3/2}}.
\end{equation}
The vector ${\bG}$ is the external force relative to the typical friction force, while $B$ measures the importance of the "angular" diffusion in momentum space caused by the random potential relative to ordinary diffusion caused by the Gaussian noise. The zero dissipation limit is $B\ra\infty$, $\bG\ra\infty$, with $|\bG|/B$ kept fixed. In the new variables, the stationary Fokker-Planck equation to be solved is
\begin{equation}\label{eq:FPstationary}
    \frac{\partial}{\partial y_i}\left((G_i- y_i)f\right)=\frac{B}{y^3}\Dp f+\Delta f.
\end{equation}
One can try solving it perturbatively in $\bG$. If $f_n,$ $n=0,1,\ldots,$ is the contribution to $f$ at order $F^n$, then it satisfies 
\begin{equation}\label{eq:recursion}
  \frac{1}{y^3}\Dp f_n+\frac{1}{B}\left(\Delta f_n+\partial_i(y_i f_n)\right)=\frac{1}{B} G_i\partial_i f_{n-1}.
\end{equation}
The initial condition for this recursion is the Maxwell-Boltzmann distribution $f_0\sim e^{-y^2/2}.$ To find $f_1$ one needs to solve a rather non-trivial ODE. However, in the limit of weak dissipation the ODE reduces to an algebraic equation, and for a fixed $G/B$ one can find $f_1$ as a power series expansion in $1/B$. For $d=3$ we get
\begin{equation}
    f_1=\frac{G \cos\theta}{2B} f_0\left(y^4+\frac{1}{B}(-2y^7+9y^5)+O(1/B^2)\right).
\end{equation}
The leading term agrees with eq. (\ref{eq:f1nondiss}). The corresponding conductivity is
\begin{equation}\label{eq:conddiss}
    \sigma=\frac{16 T^{3/2}}{W\sqrt {2\pi}}-\frac{315 \nu T^3}{2 W^2}+O(\nu^2).
\end{equation}
As might be expected, dissipation reduces conductivity (it can be thought of as modeling inelastic scattering). Let us compare this result with the Matthiessen rule. For $W=0$ the conductivity would be $\sigma_{inel}=1/\nu$. Therefore the Matthiessen rule predicts that for small $\nu$ the conductivity is
\begin{equation}\label{eq:Matt}
    \sigma_{Matt}=\frac{16 T^{3/2}}{W\sqrt {2\pi}}-\frac{256 \nu T^3}{2\pi W^2}+O(\nu^2).
\end{equation}
The second term in eq. (\ref{eq:Matt}) has the right functional dependence, but the numerical coefficient is about four times smaller than the second term in eq. (\ref{eq:conddiss}).

Using eq. (\ref{eq:recursion}) one can also compute $f_2$, $f_3$, etc. for small dissipation. We only quote the leading terms in the $1/B$ expansion of $f_2$ for $d=3$:
\begin{equation}
f_2=\frac{G^2}{B^2} f_0 \left(\frac{B}{30} y^5 +O(1)\right).
\end{equation}
When expressed in terms of the original variables, this is of order $E^2/\nu$ and thus diverges when $\nu$ is taken to zero and $E$ is kept fixed. As discussed above, this is a reflection of the fact that a well-defined NESS requires nonzero dissipation. While $f_2$ does not contribute to the current for symmetry reasons, it contributes to the average kinetic energy which is given by 
\begin{equation}\label{eq:Joule}
   \int\frac12\bp^2 (f_0+f_1+f_2+\ldots) d^3p= \frac{3T}{2}\left(1+\frac{16}{3\sqrt {2\pi}}\frac{E^2\sqrt T}{\nu W}+O(E^4)\right). 
\end{equation}
The second term in parentheses represents Joule heating. It diverges in the limit $\nu\ra 0$ and fixed $E$. For LRT to apply, Joule heating must be negligible, that is, $E\ll (\nu W)^{1/2} T^{-1/4}$. For weak dissipation this is much smaller than the naive condition $E\ll W/T$. Similarly, one can find $f_3$ which contributes a term of order $E^3$ to the current. Requiring it to be negligible compared to the LRT result, we get the same constraint on $E$ as above. 

\section{NESS and variational principles}

From the Bayesian perspective \cite{Jaynes}, variational principles in statistical mechanics are best guesses for a probability distribution when only partial information about the dynamics of the system is available. For example, the Gibbs distribution is the best guess for a state of system which has been in a contact with a large reservoir, and the precise interaction between the system and the reservoir is unknown. Most textbook derivations of the Gibbs distribution are based on the MAXENT principle, which instructs us to maximize the entropy of the  probability distribution given all available information. In the case of our model, it is natural to assume that the available information is the average kinetic energy (determined by the temperature of the reservoir if the Joule heating is small) and the average current $\bj$. Since the potential $U$ is random, the precise Hamiltonian of the system is unknown, and neither is the interaction between the system and the reservoir. Applying MAXENT we get that $f(\bp)$ is a shifted Maxwell-Boltzmann distribution: $f(\bp)=f_0(\bp-\bj)\sim \exp(-(\bp-\bj)^2/2T)$. To linear order in $\bj$ this gives $f=f_0 (1+  (\bj\cdot \bp)/T+O(\bj^2)).$ The same functional form is obtained by maximizing entropy while imposing the condition that total dissipated power $\nu\langle \bp^2\rangle$ is equal to $\bE\cdot\langle\bp\rangle$, the work per unit time supplied by the drive. The MAXENT distribution disagrees with the functional form of $f_1(\bp)$ we found in our model, eq. (\ref{eq:f1nondiss}). 

Another popular variation principle for finding NESS is Minimum Entropy Production (MINEP) \cite{Prigogine,deGrootMazur,KleinMeijer}. MINEP instructs us to minimize {\it internal} entropy production in the system given available macroscopic constraints. Here one needs to distinguish internal entropy production (which is always non-negative and is strictly positive in a NESS) from the total rate of change of entropy of the system which also includes the entropy outflow to the environment. In a NESS, the total entropy production is zero by definition. Unlike entropy, which is an information-theoretic quantity, entropy production depends on dynamics. In our model, a coarsened description of dynamics neglecting the interaction with the reservoir and averaged over the random potential is given by the Fokker-Planck equation (\ref{eq:two}), and the corresponding internal entropy production is (\ref{eq:dsdt}). Extremizing it with respect to $f$ while keeping fixed the average current and normalization gives
\begin{equation}
    -2\frac{W}{p^3}\Dp f+\frac{W}{p^3}\frac{\nabla_\Omega f\cdot \nabla_\Omega f}{f}+(\lambda +\be \cdot\bp) f=0,
\end{equation}
where $\lambda$ and $\be$ are Lagrange multipliers. 
We can solve this equation perturbatively in the average current (which is proportional to the Lagrange multiplier $\be$). If we write $f=f_0+f_1+f_2+\ldots$, where $f_n$ is of order $n$ in $\be$ and $f_0$ is the Maxwell-Boltzmann distribution, then $f_1$ satisfies
\begin{equation}
    \frac{W}{p^3}\Dp f_1=\frac{1}{2} (\be\cdot\bp) f_0.
\end{equation}
Assuming $f_0$ has the Maxwell-Boltzmann form, this is the same as eq. (\ref{eq:recursion}) for $n=1$ and zero dissipation. Thus, to leading order in the current MINEP gives the correct distribution function,  in an agreement with the general arguments in Ref. \cite{Brussel} which apply to any master equation in a linear regime. However, problems appear at quadratic order, because the equation for $f_2$ does not have a unique solution. This happens because adding to $f$ any spherically-symmetric function of $\bp$ does not affect entropy production at this order. The inability of MINEP to determine $f_2$ is not surprising, since at quadratic order in the electric field the distribution function depends on the coupling to the environment, and MINEP is ignorant about it.

\section{Conclusions}

We have shown that at quadratic order in the electric field, the NESS in our model  has a singular dependence on the coupling to the environment $\nu$ (formally, it diverges as $\nu\ra 0$). Thus, the limits of zero electric field and zero $\nu$ do not commute, and LRT has a nonzero range of validity only for an open system. This supports and sharpens van Kampen's objection to LRT \cite{vankampenobjection}: it is the coupling to the environment which leads to macroscopic linearity. This coupling leads to a randomization whose effect at long times is similar to Boltzmann's {\it Stosszahlansatz}. 

Another implication is that nonlinear transport necessarily depends on the details of such a coupling and cannot be computed by extending the usual manipulations of LRT to next order in the electric field.

In our model, a coupling to the environment is required to remove the energy supplied by the electric field. Another way to create a current-carrying NESS is to couple the system to a pair of reservoirs with different chemical potentials. In this case no work is done in the bulk of the system, so no bulk dissipative coupling is needed. However, at quadratic order in the drive the NESS will depend on the Joule heating in the reservoirs. If one wants to describe a NESS at a non-linear level, one cannot avoid modeling the coupling to the environment, either in the bulk or in the reservoirs.

We also saw that MAXENT does not predict the correct NESS even at linear order in the electric field, while MINEP gets the linearized NESS right but fails at quadratic order. This is not surprising, since nonlinear corrections must diverge in the limit of zero dissipation, but MINEP does not explicitly introduce any dissipative couplings.

It would be interesting to find a non-perturbative solution to eq. (\ref{eq:FPstationary}), i.e. not to  expand in powers of $\bG/B$ (but still assume that $\bG$ and $B$ are large). It would describe a strongly non-equilibrium steady state driven by a large \footnote{That is, large enough to change the energy of the particle by a relative amount of order $1$ over the mean free time.} electric field  and stabilized by a small dissipation. It does not seem possible to find such a solution analytically, so one must resort to numerical methods.

\section*{Appendix: Derivation of the disorder-averaged Fokker-Planck equation}

Consider a Liouville-Fokker-Planck equation:
\begin{equation}\label{eq:LiouFP}
    \eps^2\frac{\partial \CF}{\partial t}+\frac{\partial}{\partial x_i}\left(p_i \CF\right)+\frac{\partial}{\partial p_i}\left((\eps^2 (E_i-\nu p_i)+\eps K_i) \CF\right)=\eps^2\nu T \frac{\partial^2\CF}{\partial p_i\partial p_i},
\end{equation}
where $E_i$ is a non-random force (possibly dependent on $t,\bx,\bp$), $-\nu p_i$ is a friction force, and $K_i(\bx)$ is a random static force which may depend on coordinates, but not the momentum of the particle. We also assume $K_i(\bx)$ has zero average. We write $\CF=\sum_{n=0}^\infty \eps^n \CF_n$ and plug this into the equation. This gives a set of coupled equations for $\CF_n$. To get a set of equations which closes we need to truncate at some order by neglecting $\CF_n$ for some $n$. We will truncate at the first nontrivial order, namely, $n=2$. The $O(\eps^0)$ equation reads:
\begin{equation}\label{eq:CF0}
    p_i\frac{\partial \CF_0}{\partial x_i}=0.
\end{equation}
The simplest way to satisfy this equation is to assume that $\CF_0$ does not depend on $\bx$. This is what Ref. \cite{KestenP} assumes. We will not make this assumption yet. 

Note that the corresponding inhomogeneous equation
\begin{equation}
    p_i\frac{\partial u}{\partial x^i}=v(t,\bx,\bp)
\end{equation}
has a formal solution
\begin{equation}
    u(t,\bx,\bp)=-\int_0^\infty v(t,\bx+\bp s,\bp)ds
\end{equation}
This formal solution becomes an actual solution if $v$ decays sufficiently rapidly  for large $|\bx|$.

The $O(\eps^1)$ equation reads:
\begin{equation}
    p_i\frac{\partial \CF_1}{\partial x_i}+K_i\frac{\partial\CF_0}{\partial p_i}=0.
\end{equation}
It has a formal solution 
\begin{equation}\label{eq:CF1}
    \CF_1(t,\bx,\bp)=\int_0^\infty \left.\left(\frac{\partial\CF_0}{\partial p_i} K_i\right)\right|_{\bx\mapsto\bx+\bp s} ds.
\end{equation}
This formal solution becomes an actual solution if $\bG$ is nonzero only in a bounded region. The general solution differs from this particular one by a solution of the homogeneous equation. We absorb it into $\CF_0$, thus $\CF_1$ is given by eq. (\ref{eq:CF1}). 

The $O(\eps^2)$ equation reads:
\begin{equation}
    \frac{\partial \CF_0}{\partial t}+\frac{\partial}{\partial x_i}\left(p_i \CF_2\right)+\frac{\partial}{\partial p_i}\left((E_i-\nu p_i)\CF_0+ K_i \CF_1\right)=\nu T \frac{\partial^2\CF_0}{\partial p_i\partial p_i},
\end{equation}
Now we average over the random force $\bK$ with $\CF_1$ given by eq. (\ref{eq:CF1}). To get a closed equation for $f=\langle \CF_0\rangle$ we neglect the term containing $\langle \CF_2\rangle$.  We also make the usual "decoupling" assumption that the triple correlator $\langle \CF_0 K_i K_j\rangle$ factorizes into a product of $\langle \CF_0\rangle$ and $\langle K_i K_j\rangle$ \cite{vanKampenstochastic}. This gives an equation for $f$:
\begin{equation}\label{eq:CF2av}
\frac{\partial f}{\partial t}+\frac{\partial}{\partial p_i}\left((E_i-\nu p_i)f\right)=\nu T \frac{\partial^2 f}{\partial p_i\partial p_i}+\frac{\partial Q_i}{\partial p_i} ,
\end{equation}
where
\begin{equation}
    Q_i(\bx,\bp)=\int_0^\infty \frac{\partial f(t,\bx+\bp s,\bp)}{\partial p_k}\langle K_i(\bx) K_k(\bx+\bp s)\rangle ds.
\end{equation}
This has to be solved along with the average of eq. (\ref{eq:CF0}):
\begin{equation}\label{eq:CF0av}
    p_i\frac{\partial f}{\partial x_i}=0.
\end{equation}

Eq. (\ref{eq:CF2av}) is non-local in space. Let us look for solutions  $f(t,\bx,\bp)$ which not depend on $\bx$, which is appropriate if $E_i$ is spatially uniform and $\bG$ is stationary in space. Then eq. (\ref{eq:CF0av}) is satisfied, while eq. (\ref{eq:CF2av}) becomes a Fokker-Planck equation:
\begin{equation}\label{eq:FPav}
    \frac{\partial f}{\partial t}+\frac{\partial}{\partial p_i}\left((E_i-\nu p_i)f\right)=\nu T \frac{\partial^2 f}{\partial p_i\partial p_i}+\frac{\partial}{\partial p_i}\left(w_{ik}(\bp)\frac{\partial f}{\partial p_k}\right),
\end{equation}
 where 
 \begin{equation}
     w_{ik}(\bp)=\int_0^\infty \langle K_i(0) K_k(\bp s)\rangle ds.
 \end{equation}

To simplify further, suppose $\bK=-\nabla U$, where $U(\bx)$ is a random potential such that $\langle U(\bx) U(\by)\rangle = C(|\bx-\by|)$ depends only on $r=|\bx-\by|$. Then
\begin{equation}
    \langle K_i(0) K_j(\bx)\rangle=-\left(\delta_{ij}-\frac{x_i x_j}{r^2}\right) \frac{C'}{r}-\frac{x_ix_j}{r^2} C''.
\end{equation}
Assuming that $C'(0)=0$ (which has to be the case if $\langle U(\bx) U(\by)\rangle$ is a smooth function of $\bx-\by$), the contribution of the second term on the r.h.s. to $w_{ik}(\bp)$ is zero, and we get
\begin{equation}
    w_{ik}(\bp)=\left(\delta_{ik}-\frac{p_i p_k}{p^2}\right)\frac{W}{p},
\end{equation}
where $W=-\int_0^\infty C'(r) \frac{dr}{r}$. Thus, for a random potential force eq. (\ref{eq:FPav}) reduces to eq. (\ref{eq:two}).

\bibliography{Bibliography}

\end{document}